# Efficient Kernel Fusion Techniques for Massive Video Data Analysis on GPGPUs

Asif M. Adnan and Sridhar Radhakrishnan
*School of Computer Science*
*University of Oklahoma*
*Norman, USA*
*{asifmadnan,sridhar}@ou.edu*

Suleyman Karabuk
*School of Industrial Engineering*
*University of Oklahoma*
*Norman, USA*
*karabuk@ou.edu*

*Abstract*—Kernels are executable code segments and kernel fusion is a technique for combing the segments in a coherent manner to improve execution time. For the first time, we have developed a technique to fuse image processing kernels to be executed on GPGPUs for improving execution time and total throughput (amount of data processed in unit time). We have applied our techniques for feature tracking on video images captured by a high speed digital video camera where the number of frames captured varies between 600-1000 frames per second.

Image processing kernels are composed of multiple simple kernels, which executes on the input image in a given sequence. A set of kernels that can be fused together forms a partition (or fused kernel). Given a set of Kernels and the data dependencies between them, it is difficult to determine the partitions of kernels such that the total performance is maximized (execution time and throughput). We have developed and implemented an optimization model to find such a partition.

We also developed an algorithm to fuse multiple kernels based on their data dependencies. Additionally, to further improve performance on GPGPU systems, we have provided methods to distribute data and threads to processors. Our model was able to reduce data traffic, which resulted better performance. The performance (both execution time and throughput) of the proposed method for kernel fusing and its subsequent execution is shown to be 2 to 3 times higher than executing kernels in sequence. We have demonstrated our technique for facial feature tracking with applications to Neuroscience.

*Keywords*-kernel fusion, GPU, parallel image processing, tracking, video analysis

## I. INTRODUCTION

Application of High Speed Digital Video (HSDV) for identifying previously unseen patterns has become very popular. HSDV and its analysis have applications in industrial product development and quality control [1], Neuroscience [2], and others [3]. Ross et. al [2] used HSDV for identifying different traits of human facial action movements. One of their distinctive findings is that facial muscles on the left side (controlled by right hemisphere of brain) start moving earlier than the right side, for spontaneous facial expressions. Several such discoveries can be made possible by analyzing the data produced by HSDV.

HSDV results in massive volume of data to store and demands extensive computing power to analyze the data. A one second long video image captured at 600 frames per second (fps) (resp. 1000 fps) with frame dimensions $192 \times 432$ pixels (resp. $800 \times 600$ pixels), requires about 190MB (resp. 1.8GB) of storage. Clearly, processing HSDV data using CPUs will be challenging because of its tremendous volume. Real-time or near real-time processing would be further challenging because of high volume data transfer requirement through the computation components (for example, disk to main memory).

In this paper, the term "*Image Processing Algorithm*" or *IPA* will be used to refer to both image and video processing algorithms. A complex IPA, *Alg*, can be viewed as a sequence of multiple simple algorithms, $A = \{A_i : 1 \leq i \leq n\}$. The algorithms have to be executed in a sequence on the entire video data $I$ having dimensions $N \times M \times T$. Here, $(N \times M)$ refers to image spatial dimension or frame dimensions, and $T$ refers to temporal dimension or number of frames.

A general CUDA implementation of *Alg* will be to develop kernels $K_i$ for each of the algorithms, $A_i$, and execute them in sequence. In this paper, we will use the terms algorithms and kernels interchangiably. Each of the $K_i$ will be distributed among the Streaming Multiprocessors (SMs) and results in the intermediate output. Intermediate output $I_{out_i}$ generated by $K_i$ will be used by the next kernel $K_{i+1}$ in the sequence. Please note a particular GPU device will have a number of SMs and each SMs have a set number of processors and local memory (called a shared memory). Additionally, each of the SMs and its processor has access to global memory. There are no concurrent accesses to global memory, while there are concurrent read on shared memory by processors within a SM. Thus, keeping as much of data as possible closer to the processors (e.g. shared memory) is better for faster data access.

Image processing algorithms are memory access intensive. A simple sequential execution of kernels $K_1, ..., K_n$ will result in increased data traffic among different GPU memory units (global and shared), as well as, it will use global memory to store generated intermediate data. The overall performance is related with the volume of data traffic between global and shared memories. Performance will increase with the decrease of data traffic and reduction in global memory usage.

The overall structure of our technique to improve execution time on a set of image processing kernels is as follows. First, we will determine the data access patterns of the given set of kernels. The data access patterns will provide information on the amount of data that needs to be transferred from global memory to shared memory for execution. Second, based on the data access patterns of each of the kernels, we will partition the set of kernels, where each partition will result in a fused kernel. The set of fused kernels are executed in a sequence consistent with the sequential execution of the set of kernels. Among all possible partitions, our goal it to determine a partition that will reduce the overall data traffic between global memory and shared memory. In order to determine this partition, we have developed a optimization model and have used the Gurobi [4] tool to implement the model. We also would like to point out that once a partition is determined, we have to provide a mechanism to perform the fusing. We have provided one such technique in this paper. Third, once the fused kernel is formed, we will provide techniques to allocate resources (threads, processors, shared memory, global memory) optimally to reduce the total execution time of *all* fused kernels.

In section II we presented a brief overview of CUDA enabled GPU architecture. Typical image processing algorithm and its implementation on GPU is presented in section III. In section IV we have discussed the data access patterns that are common in image processing applications. Once the threads have been allocated for executing the kernels, there will be data dependencies among the threads and these are discussed in section V. We presented our optimal kernel fusion model and in section VI. In the same section we provided a mechanism to perform the fusing once the set of kernels chosen to be fused have been determined. We also showed how the performance of a fused kernel can be improved by efficient resource allocation and data distribution. Experiments and analysis of the proposed techniques for facial feature tracking is discussed in sections VII and VIII respectively. We discuss this work and compare with the other kernel fusion techniques in section IX and conclude in section X.

## II. CUDA Architecture Overview

GPU architecture generally consist of multiple Streaming Multiprocessors (SM), different levels of memory units. Each SM has multiple processors (cores), local memory unit, such as Shared Memory (SHMEM), and registers. Computation heavy and parallelizable tasks are transferred from CPU to GPU, and executed as set of instructions.

CUDA provides a general-purpose parallel programming model which extends C and defines C-like functions, called kernels. Kernels are executed in parallel by different threads on the GPU. Threads are grouped together in blocks, which can be either one, two or three dimensional. All the blocks of threads, form a one, two or three dimensional grid. Threads within a block can co-operate among themselves through the SHMEM and synchronize their execution to coordinate memory accesses. Threads are managed and executed by the multiprocessor in groups. This group is called *warp*. The usual warp size is 32 threads.

Data communication between CPU and GPU are conducted through the Global Memory (GMEM). All threads in a SM operate in a Single Instruction Multiple Data (SIMD) fashion, while threads belonging to two or more SMs operate in a Single Instruction Multiple Thread (SIMT) fashion. Data from GMEM is accessed by threads for computation. The result of the computation is written back to the GMEM as dictated by the algorithm. SHMEM access is a couple of magnitude faster than GMEM access. Before any algorithm executes, it is desirable to bring the data to SHMEM. The limited space of the SHMEM is a bottleneck to this approach. A more detailed description of CUDA enabled GPUs can be found in [5].

## III. Image Processing Kernels

In this section we will present a general image and video analysis application using CUDA enabled GPU devices. A typical feature tracking IPA on a set video frames consist of the application multiple kernels. These include kernels for reducing noise, enhancing pixel quality, image transformation say from RGB to Gray-scale, identifying features (using say Gaussian Filter and Gradient Smoothing), and finally tracking them on video data. Noise reduction and pixel enhancement kernels takes as input raw image data and generates better quality data. The output of this stage is used by other kernels (image transformation, Gaussian filter, gradient smoothing filer, and others) to detect the features. Finally different particle filter (such as the Kalman filter) algorithms are applied to track the features in different frames of the video data.

We modeled a $T$ second long video having $N \times M$ dimensional frames, as, $I[d_x, d_y, d_t]$, where, $d_x = N$, $d_y = M$, and $d_t = T$. Video segments also have frame rate represented as $R$ frames per second. A $T$ second long frame with $R$ frames per second has total, $F = T \times R$ frames. A pixel of a frame $I[d_x, d_y, t]$ at location $(i, j)$, will be represented as $I[i, j, t]$.

Lets say that a complex algorithm $Alg$ consists of a set of algorithms, $A = \{A_i : 1 \leq i \leq n\}$. A very general implementation of $Alg$ on CUDA devices would be to develop single kernels $K_i$ for each of the algorithms $A_i$. When the kernels $K_i$ $K = \{K_i : 1 \leq i \leq n\}$ finishes execution, the complex $Alg$ completes execution. Fig 1 is shows multiple kernels in sequence and along with their data dependencies.

For a sequence of kernel executions, $K_1, K_2, ..., K_n$, processing $I_{in}[d_x, d_y, d_t]$, we will get,
$I_{out_{K_1}}[d_x, d_y, d_t] = K_1(I_{in}[d_x, d_y, d_t])$
$I_{out_{K_2}}[d_x, d_y, d_t] = K_2(I_{out_{K_1}}[d_x, d_y, d_t])$
...
...
$I_{out_{final}}[d_x, d_y, d_t] = K_n(I_{out_{K_{n-1}}}[d_x, d_y, d_t])$
So, the final output will be generated after executing a sequence of kernels on the given input data, as in,
$I_{out_{final}}[d_x, d_y, d_t] = K_n(...(K_1(I_{in}[d_x, d_y, d_t])))$.

## IV. Data Access Patterns

Different algorithms have different data access patterns, for example, consider the algorithms mentioned in the previous section. The algorithms for noise reduction and pixel enhancement algorithms relies both on spatial and temporal neighbors of a pixel. Image transformation algorithms such as from RGB to Gray scale transformation relies on a single pixel. Please note that there exists other transformation algorithms that requires spatial and/or temporal neighbor values. Gaussian filter and gradient smoothing algorithms that we used for our application are spatial filters, and hence, they require spatial neighbors. The Kalman filter algorithm uses the identified feature position in multiple frames, for tracking. Such a filter relies on single point in the spatial dimension, but also requires the temporal neighbors.

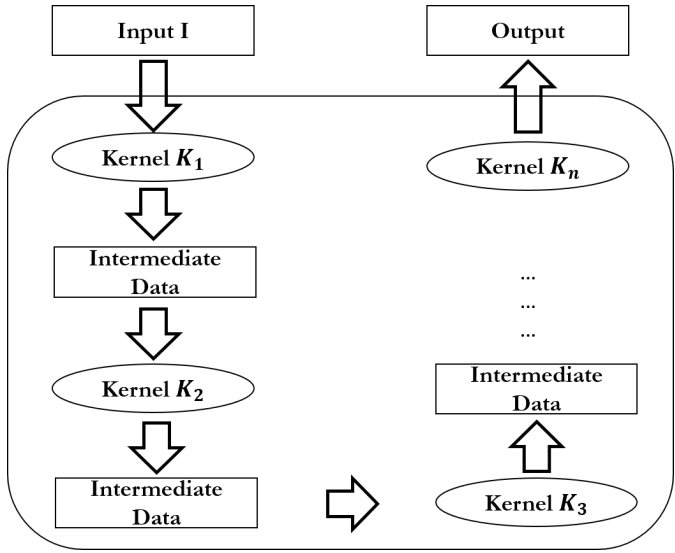

Figure 1: Kernel Data Dependency Diagram. Kernels $K_1, ..., K_n$ are executed in sequence. Kernel $K_{i+i}$ takes in as input the output $I_{out\,K_i}$ generated by kernel $K_i$.

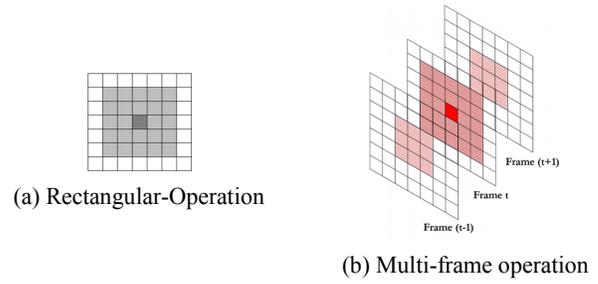

(a) Rectangular-Operation

(b) Multi-frame operation

Figure 2: $I_{out}[i, j, t] = Function(I_{in}[d_i^0, d_j^0, d_t^0])$

Table II: Image Processing Steps and Types

| Algorithms | Type of Operation | Multi-Frame |
|---|---|---|
| Convert RGBA to Gray | Point Operation | No |
| IIR Filter | Point Operation | Yes |
| Gaussian Smooth Filter | Rectangular Operation | No |
| Gradient Filer | Rectangular Operation | No |
| Threshold Computation | Rectangular Operation | No |
| Apply Kalman Filter | Single Point Operation | Yes |

Computing the value of a pixel $I_{out}[i, j, t]$ involves a function $Function(I_{in}[d_i^0, d_j^0, d_t^0])$. Here, $I_{out}[i, j, t]$ represent a single pixel, and $I_{in}[d_i^0, d_j^0, d_t^0]$ represent the pixels neighboring to $I_{in}[i, j, t]$. The function $Function(I_{in}[d_i^0, d_j^0, d_t^0])$ generates the output pixel with $I_{in}[d_i^0, d_j^0, d_t^0]$ as input.

To generate an output box (denoted $Box_b$) of dimension $x \times y \times t$, the input box (denoted $Box_{b_{in}}$) will have dimension $(x + \delta_x) \times (y + \delta_y) \times (t + \delta_t)$, where $\delta$ is the small increment in size (the larger box).

The output value of a single pixel $I_{out}[i, j, t]$ depends on its spatial, temporal, or both spatio-temporal neighbors (Fig. 2). After analyzing commonly used image processing algorithms, we divided them into different types. Table I lists the operation types and shows their data dependency criteria. Table II classifies different simple and basic algorithms into different types based on type of operation and if it involves single or multiple frames (as in the case of video data).

Table I: Different Types of Operations

| Types of Operations | Data Dependency |
|---|---|
| Single-Point Operation | $|d_i^0| = 1, |d_j^0| = 1, |d_t^0| = 1$ |
| Rectangular Operation | $|d_i^0| > 1, |d_j^0| > 1, |d_t^0| = 1$ |
| Single-Frame Operation | $|d_t^0| = 1$ |
| Multi-Frame Operation | $|d_t^0| > 1$ |
| Spatio-Temporal Operation | $|d_i^0| > 1, |d_j^0| > 1, |d_t^0| > 1$ |

V. THREAD DEPENDENCY AND KERNEL FUSION

Each of the kernels, $K_i$ are executed as a grid of blocks, where each of the blocks will have a set of threads. The total execution time for executing the kernels $K_i \in K$ corresponding to $Alg$ will be,

$$Total^T = \sum_{i=1}^{n} (T_{Access}^{T_i} + T_{Compute}^{T_i} + T_{Write}^{T_i}) \quad (1)$$

where, for each kernel $K_i$, $T_{Access}^{T_i}$ is the data access time from GMEM to thread's local memory, $T_{Compute}^{T_i}$ is the time to compute accessed data, and $T_{Write}^{T_i}$ represents the time to write/store data to GMEM.

Each of the thread block will generate output having certain dimensions. Fig. 3 is showing the division of the video data of dimension $N \times M \times T$ into boxes $Box_b$ of dimension $x \times y \times t$. There will be $B = \frac{N \times M \times T}{x \times y \times t}$ number of such boxes. As each thread block will perform computation on the data in each of the boxes. Without loss of generality we can assume that there will be $B$ number of thread blocks. All these thread blocks will be distributed among $\rho_{SM}$ number of SMs and executed via warps (or thread groups). To generate an output of each box $Box_b$, each thread block of the kernels has to take input a box $Box_{b_{in}}$.

A. Thread Dependency Types

Data access patterns discussed in the previous sub-section leads to different types of thread and block-level thread dependencies. More the dependencies less will be the achievable parallelism. In our analysis we found three basic dependency types:

a) Thread to Thread (TT): This occurs when a thread $T[t_x, t_y, t_z]_i$ of kernel $K_i$ depends on the output of the corresponding $T[t_x, t_y, t_z]_{i-1}$ of kernel $K_{i-1}$. Every thread in here requires output from the corresponding thread to complete its execution. This results in

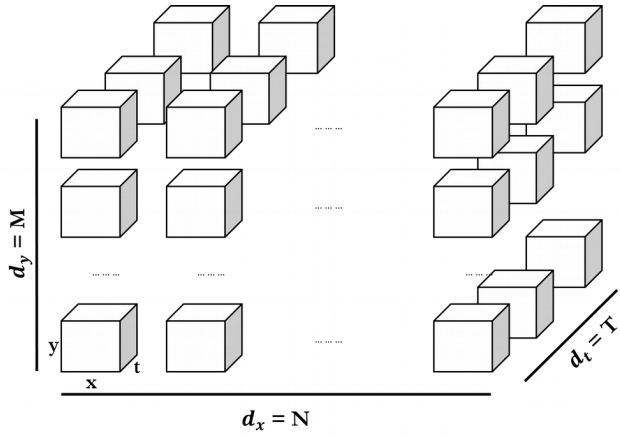

Figure 3: Thread block $TB_b$ generates output box $Box_b$. There will $B = \frac{N \times M \times T}{x \times y \times t}$ number of such boxes. Correspondingly, there will be $B = \frac{N \times M \times T}{x \times y \times t}$ numbers of thread blocks per kernel $K_i$. Each of the thread blocks $TB_b$ will compute the output $Box_b$ while the input will be $Box_{b_{in}}$

high level of parallelism or the smallest amount of dependencies. This is shown in Figure 4(a).

b) Thread to Multi-Thread (TMT): When a thread $T[t_x, t_y, t_z]_i$ of kernel $K_i$ depends on multiple threads of kernel $K_{i-1}$, we say that TMT dependency occurs. A thread has to only wait for the result of the corresponding thread block to complete its execution. Since all blocks are executed in parallel, the dependency is somewhat lower.

c) Kernel to Kernel (KK): When a block of threads that is executing a kernel $K_i$ depends on the output of multiple blocks of threads executing kernel $K_{i-1}$, we define it as KK dependency. This is shown in Figure 4(b) As an example, consider the two kernels Center detection to identify the feature and Kalman filter kernel to track features. These two kernels have to be executed in sequence resulting in lower parallelism or higher dependencies. This is shown in Figure 4(c).

*B. Kernel Fusion*

Kernel fusion is a source code transformation technique, where new kernels are created by aggregating or merging codes of multiple other kernels. Usually kernels executing on same data are good candidates for fusion.

In case of kernel fusion, a fused kernel's total execution time will be (in contrast with equation 1), as follows:

$$Total^T = T^{T_1}_{Access} + \sum_{i=1}^{n} (T^{T_i}_{Compute}) + T^{T_n}_{Write} \quad (2)$$

Here, we considered that, for a given set of kernels, $K$, having $n$ number of kernels, the necessary input box $Box_{b_{in}}$ will be copied from GMEM to SHMEM. Now all the fused kernels will access data from SHMEM and store intermediate data in SHMEM. After the execution of the last kernel in the sequence, the final result will be stored in GMEM.

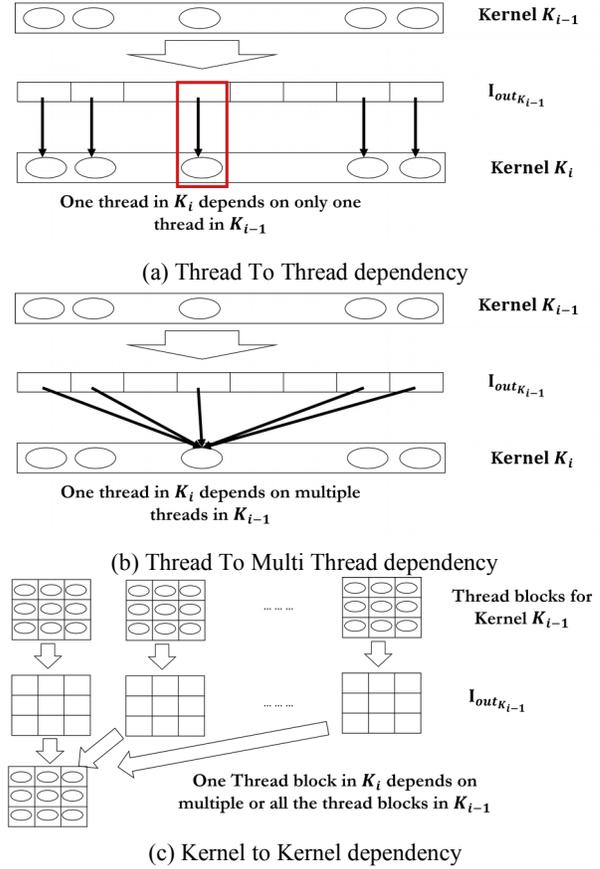

(a) Thread To Thread dependency

(b) Thread To Multi Thread dependency

(c) Kernel to Kernel dependency

Figure 4: Thread and Block level Data Dependency Types

Fusing multiple kernel reduces total number of data transfers among GPU memory units. Reduction in data transfer to and from GMEM and local memory units also reduces data access from more time consuming global memory access. As the kernels $K_1, ...K_n$ are generating and reusing intermediate data to and from SHMEM or local resources, GMEM is freed up for additional data to be loaded from CPU's main memory.

## VI. OPTIMAL KERNEL FUSION

Wahib and Maruyama [6] described a technique to estimate the execution time for kernels with heavy data access on GPU systems. We will use this estimate to determine the optimal partition of a set of kernels given in sequence for execution. Each partition will be the fused kernel. For example, consider three kernels that needs to be executed in sequence as follows: $K_1, K_2, K_3$. The possible partitions are $(\{K_1\}, \{K_2\}, \{K_3\}), (\{K_1, K_2, K_3\}), (\{K_1, K_2\}, \{K_3\}), (\{K_1\}, \{K_2, K_3\})$. The partition with a minimum total execution time is selected to be the optimal one, where the execution time is determined using the technique in [6].

*A. Identify Fusable Kernel Sets*

For a target task, we decompose the task into multiple kernels. After analyzing the thread and block level dependencies, we generate multiple sets, $K_1, K_2, ..., K_m$ of fusable kernels, where, $K_k = \{K_a : 1 \leq a \leq n_k\}$. For generating possible fusable kernel sets, we considered kernels having only Thread to Thread and Thread to Multi-Thread dependency on the previous kernel(or fused kernel) in the sequence. We excluded KK dependent kernels. We developed the following model for kernel fusion, for each set of fusable kernels and was solved using the Gurobi

---

Number of Algorithms in a fusible set, $K_k \{A_1, ..., A_j\}$: $n$

Number of Possible fused kernel combinations, $\{K_1, ..., K_i\} : i = \frac{n(n+1)}{2}$

$\vec{a_i}$ is a vector, having $j$ number of elements, $\begin{pmatrix} 0 \\ 1 \\ 1 \\ 0 \\ ... \\ ... \\ 0 \end{pmatrix}$

where, $a_{i,j} = 1$, if algorithm $A_j$ is selected in kernel. $a_{i,j} = 0$, otherwise.
$X_i = \{0,1\}$; where, $X_i = 1$, if kernel $K_i$ is selected. Otherwise, $X_i = 0$.
$C_i$ is the predicted execution time for executing $K_i$.

**Objective**: minimize $\sum(X_i C_i)$.
s.t: $\sum(X_i \vec{a_i}) = 1, \forall\ j$.

---

Figure 5: Modeling the Objective Function

### B. Algorithm to fuse selected kernels

In the previous step, we generated sets of kernels to be fused. Algorithm 1 will take input as input a set of kernels and generate a fused kernel $K_f$.

---

**Algorithm 1:** Fuse kernels $K_1, K_2, ..., K_n$ into $K_f$

**Data**: Set of kernels, $K = \{K_i : 1 \leq i \leq n\}$
**Result**: Fused kernel $K_f$
1 Copy Input Box, $Box_{b_{in}}$ from Global to shared memory
2 **for** each $K_i$ **do**
3     Convert Global Memory access into Shared Memory access.
4     Insert instruction of $K_i$ into $K_f$
5     if $K_i$ and $K_{i+1}$ are Thread to Multi-Thread dependent then Insert Synchronization.
6 **end**
7 Copy/Store SHMEM data into GMEM.

---

Line 1 of Algorithm 1 copies the required input box from the GMEM to SHMEM. Conversion of GMEM access to local or SHMEM access is explained in the next sub-section. Adjustment of input and output box dimensions are very important and is explained in a subsequent sub-section.

Local synchronization primitives are inserted in line 5 that is based on thread to multi-thread dependencies that are observed. Table III is showing two different types of fusion for three sample kernels. Here, the kernels are synthetic kernels, but has are very close to the ones used in our experiments.

Table III: Simple and Fused Kernel Samples

```
__global__ RGB2Gray(Iin, Iout)
{
  int i = blockIdx.x blockDim.x +
    threadIdx.x;
  int threadID = blockIdx.y blockDim.y +
    threadIdx.y;

  for(int ii = 0; ii< PixelPerThread; ii++)
  for(int jj = 0; jj< PixelPerThread; jj++)
    Iout[i+ii, j+jj] =
    OperationRGB(Iin[i+ii, j+jj]);
}
```

```
__global__ K-Spatial(Iin, Iout)
{
  int i = blockIdx.x blockDim.x +
    threadIdx.x;
  int threadID = blockIdx.y blockDim.y +
    threadIdx.y;

  for(int ii = 0; ii< PixelPerThread; ii++)
  for(int jj = 0; jj< PixelPerThread; jj++)
  { Iout[i+ii, j+jj] =
    OperationGaussian(Iin[i-1 to i+ii+1,
    j+jj-1 to j+jj+1]);}
}
```

```
__global__ RGBFusedTh(Iin, Iout, TH)
{
  int i = blockIdx.x blockDim.x +
    threadIdx.x;
  int threadID = blockIdx.y blockDim.y +
    threadIdx.y;

  int thx = threadIdx.x;
  int thy = threadIdx.y;

  for(int ii = 0; ii< PixelPerThread; ii++)
  for(int jj = 0; jj< PixelPerThread; jj++)
    Shared[thx+ii, thy+jj] =
            Iin[i+ii, j+jj];
  __syncthreads();

  for(int ii = 0; ii< PixelPerThread; ii++)
  for(int jj = 0; jj< PixelPerThread; jj++)
  { Shared[thx+ii, thy+jj] =
    OperationRGB(Shared[thx+ii, thy+jj]);

    if(Shared[thx+ii, thy+jj]>=TH)
    { Shared[thx+ii, thy+jj] = WHITE;}
    else Shared[thx+ii, thy+jj] = BLACK;}

  for(int ii = 0; ii< PixelPerThread; ii++)
  for(int jj = 0; jj< PixelPerThread; jj++)
    Iout[i+ii, j+jj] =
    Shared[thx+ii, thy+jj];
}
```

```
__global__ Threshold(Iin, Iout, TH)
{
  int i = blockIdx.x blockDim.x +
    threadIdx.x;
  int threadID = blockIdx.y blockDim.y +
    threadIdx.y;

  for(int ii = 0; ii< PixelPerThread; ii++)
  for(int jj = 0; jj< PixelPerThread; jj++)
    if(Iin[i+ii, j+jj] >=TH)
    { Iout[i+ii, j+jj] = WHITE;}
    else Iout[i+ii, j+jj] = BLACK;}
}
```

```
__global__ RGBFusedK-Spatial(Iin, Iout)
{
  int i = blockIdx.x blockDim.x +
    threadIdx.x;
  int threadID = blockIdx.y blockDim.y +
    threadIdx.y;

  int thx = threadIdx.x;
  int thy = threadIdx.y;

  for(int ii = 0; ii< PixelPerThread; ii++)
  for(int jj = 0; jj< PixelPerThread; jj++)
    Shared[thx+ii, thy+jj] =
            Iin[i+ii, j+jj];
  __syncthreads();

  for(int ii = 0; ii< PixelPerThread; ii++)
  for(int jj = 0; jj< PixelPerThread; jj++)
    Shared[thx+ii, thy+jj] =
    OperationRGB(Shared[thx+ii, thy+jj]);

  __syncthread();

  for(int ii = 0; ii< PixelPerThread; ii++)
  for(int jj = 0; jj< PixelPerThread; jj++)
    Shared[thx+ii, thy+jj] =
    OperationGaussian(
    Shared[thx+ii to thx+ii+1,
    thy+jj-1 to thy+jj+1]);

  for(int ii = 0; ii< PixelPerThread; ii++)
  for(int jj = 0; jj< PixelPerThread; jj++)
    Iout[i+ii, j+jj] =
    Shared[thx+ii, thy+jj];
}
```

RGB2Gray() and Threshold() are two kernels. Threshold() has thread to thread dependency on RGB2Gray(). RGBFusedTh() is the kernel generated by fusing these two kernels. The individual kernel K-Spatial() has the thread to multi-thread dependency on RGB2Gray(). When we fuse these two kernels, we get the kernel RGBFusedK-Spatial(). In both the fused kernels, we calculated the pixel index only once for each thread block to further improve performance. At the very beginning, we copied all the necessary pixels from GMEM to SHMEM. The set of instructions were inserted from corresponding kernels in order, after converting GMEM accesses to SHMEM access (more on this in the next sub-section). After the execution of all the instructions, the fused kernel copies/stores SHMEM data into GMEM.

Lets consider the case wherein the fused kernel $K_i$ will be executed by a grid of blocks (with $B = \frac{N \times M \times T}{x \times y \times t}$, and each block will be executed as a set of threads, having $Th_x \times Th_y \times Th_t$ threads. To access any GMEM element, both block-offset and thread-offset are necessary. Note that block-offset = $FunctionOf(BlockID)$ and thread-offset = $FunctionOf(ThreadID)$.

When, we convert any global access to local shared memory access, we can omit the block-offset, because all the necessary data is copied into SHMEM from the GMEM. Table III shows examples of such access conversion.

### C. Data Distribution for Executing a Fused Kernel

In the previous section, we showed that, to generate an output box $Box_b$ (by a thread block $TB_b$), it takes as input the box $Box_{b_{in}}$. For computation on a box $Box_b$ with dimensions $x \times y \times t$, the dimensions of the input box $Box_{b_{in}}$ will be $(x + \delta_x) \times (y + \delta_y) \times (t + \delta_t)$ (Fig 6). Algorithm 2 will generate the size of the input box $Box_{b_{in}}$ for a given execution sequence of $n$ kernels $K_i$.

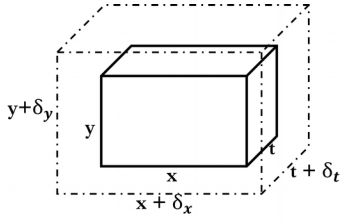

Figure 6: Input Output Data dependency for generating each Box $Box_b$. Each corresponding kernel will input a box $Box_{b_{in}}$ and generates the box $Box_b$. Dimension of $Box_b$ is $x \times y \times t$, and dimension of $Box_{b_{in}}$ is $(x + \delta_x) \times (y + \delta_y) \times (t + \delta_t)$.

---

**Algorithm 2:** Find input size for generating output image box $Box_b$

**Data**: Set of kernels, $K = \{K_i : 1 \leq i \leq n\}$
**Result**: Size of Input Box $Box_{b_{in}}$

1 Set the dimensions of $Box_{b_{in}}$ as $[x, y, t]$
2 $\delta_{x_{in}} = 0$
3 $\delta_{y_{in}} = 0$
4 $\delta_{t_{in}} = 0$
5 **for** each $K_i$ $(K_1, ..., K_n)$ **do**
6    If one thread in $K_i$ requires $(2\delta_{x_i} + 1) \times (2\delta_{y_i} + 1) \times \delta_{t_i}$ pixels,
7    If $(\delta_{x_i} \geq \delta_{x_{in}})$ $\delta_{x_{in}} = \delta_{x_i}$
8    If $(\delta_{y_i} \geq \delta_{y_{in}})$ $\delta_{y_{in}} = \delta_{y_i}$
9    If $(\delta_{t_i} \geq \delta_{t_{in}})$ $\delta_{t_{in}} = \delta_{t_i}$
10 **end**
11 The dimensions of $Box_{b_{in}}$ will be $[x + \delta_{x_{in}}, y + \delta_{y_{in}}, t + \delta_{t_{in}}]$

---

Algorithm 2 considers the data dependency parameters for all the kernels $K_1, ..., K_n$ which are fused in $K_f$ to generate the size of the desired input box $Box_{b_{in}}$. If we do not adjust the input boxes, threads computing the boundary values, will be required to access data from outside of the thread block. As CUDA devices do not allow blocks to share data with other blocks, this will lead to accesses to data GMEM. Our algorithm for computing the size of the input box allows to distribute the boxes such that, no thread has to depend on threads in other blocks.

### D. Fused Kernel Reduces Data Traffic

In this subsection, we have shown that the fused kernel has advantage over executing them in a sequence. Let $N \times M \times T$ be the size of the input that is distributed into $x \times y \times t$ dimensional boxes. Each such box, $Box_b$ is processed by a corresponding thread block, $TB_b$. The number of boxes, $B = \frac{N \times M \times T}{x \times y \times t}$. For $n$ number of kernels, $K_1, ..., K_n$, total data transfers, will be $Tranfer_{serial} = 2 \times n \times B \times x \times y \times t$.

In case of $n$ kernels fused into one kernel, number of thread blocks required is $B$. In that case, there will be total, Transfer$_{fused}$ = $2 \times B \times x \times y \times t + (x \times \delta_y + y \times \delta_x + \delta_x \times \delta_y)(t + \delta_t)$ number of transfers.

### E. Fused Kernel's Data Utilization and GPU Occupancy

Data utilization is a measure to determine the amount of space occupied in the shared memory. Since shared memory access if faster it is useful to achieve higher data utilization. The data utilization of each thread block is computed as follows:

$$DU = \frac{Output}{Input} = \frac{x \times y \times t}{(x + \delta_x) \times (y + \delta_y) \times (t + \delta_t)} \quad (3)$$

From our observation, we have found that the data utilization will be high when $x \times y \times t$ is higher. In other words, if we can use maximal SHMEM, we can achieve higher data utilization.

In order to maximize data utilization, it is required to maximize $DU_A$ in equation (3). As, $x, y$ are both referring to spatial dimensions, without loosing the generality, we can assume $x = y$. Hence,

$$DU_K = \frac{x^2 \times t}{(x + \delta_x)^2 \times (t + \delta_x)} \quad (4)$$

As, $x \times y \times t \leq \beta_{Shared}$ (where $\beta_{Shared}$ is the size of SHMEM), $x^2 \times t \leq \beta_{Shared}$. Now maximizing $DU_K$ will be equivalent to minimizing

$$V = (x + \delta_x)^2(t + \delta_t) \quad (5)$$

Solving the equation 5, we found,

$$x = y = \sqrt[3]{2\beta \frac{\delta_x}{\delta_t}}, t = \frac{1}{2^{\frac{2}{3}}} \beta^{\frac{1}{3}} (\frac{\delta_t}{\delta_x})^{2/3} \quad (6)$$

The values found in 6 will provides the minimum value for $V$. Using these values, our algorithm will decide the size of each window for distribution.

Fig 7 is showing the variation of data utilization in different CUDA devices, for different data box sizes. The limited size of SHMEM in different devices is imposing restriction on the achievable data utilization value. K20 and Gtx-750 devices has same maximum amount of SHMEM, where as C1060 allows lesser amount of maximum SHMEM. Not all combinations of data box sizes will result optimal data utilization.

An important measure that is commonly used for compute intensive GPU tasks is GPU occupancy. A high GPU occupancy indicates a high level of parallelism. More formally, GPU occupancy is the ratio of the number of thread blocks (or wraps) and maximum number of thread blocks (or wraps) per streaming multiprocessors. Increasing

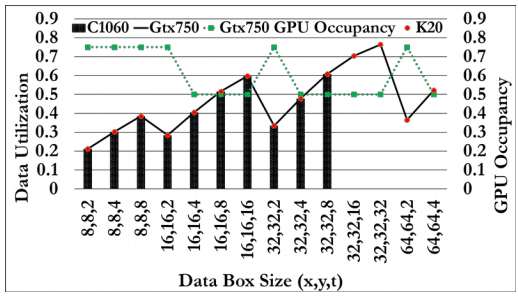

Figure 7: Data Utilization for different box size for Different Devices. Devices have limited size of SHMEM. Zero data utilization for a box with dimensions $[x, y, t]$ implies that $(x \times y \times t) >$ the size of SHMEM.

GPU occupancy will result in reduced SHMEM space for each of the blocks. Our video analysis algorithms are less compute intensive and requires more memory accesses. It is advantageous to keep have more SHMEM per block and hence we have to reduce GPU occupancy.

## VII. EXPERIMENTAL SETUP

### A. Data set we used

Ross et. al [2] used HSDV to identify facial actions in posed and spontaneous emotional expressions. We used the same data set for our experiments and implemented parallel CUDA kernel mimicking their manual tracking process. Fig. 8 is showing a frame from one of the samples [1]. Fig. 8b is showing the selected rectangles containing the target objects (external markers).

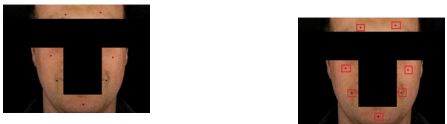

(a) Frame with external markers
(b) Interest area marked

Figure 8: Sample Frame of Tracking Facial Actions in High Speed Video

The frame rate of those videos varied from 600 to 1000 frames per second. The original frame dimension was $432 \times 192$ pixel. For performing experiments with different spatial dimensions, we preprocessed the videos and converted the dimensions into $256 \times 256$, $512 \times 512$ and $1024 \times 1024$ pixels. In case of temporal dimension, we considered computing for 1000 frames.

Our experimental setup consists of three different CUDA devices. The list is, a) Tesla C1060, b) Tesla K20 and c) Gtx 750 Ti. These devices represent two generations of Nvidia architectures. Tesla C0160 and Tesla K20 are from Tesla architecture and Gtx750 Ti is from Maxwell architecture.

We developed individual kernels ($K_1, K_2, K_3, K_4, K_5, K_6$) for each of the algorithms presented in Table II. The algorithms (and hence the kernels) will be executed in the given order in the table.

[1] Eye, nose and mouth areas are obliqued purposefully for this presentation

Table IV is showing the corresponding kernel names and relevant thread level dependencies for each kernel. We will refer each of these kernels as "Simple Kernel". Execution of these kernels in sequence will be referred as "No Fusion". From the dependency level, we created two fusable kernel sets, $\mathsf{K}_1$ consisting of $K_1, K_2, K_3, K_4, K_5$ and $\mathsf{K}_2$ consisting of $K_6$.

By using the process described in section VI-A, we found a fusion solution to fuse all the kernels in $\mathsf{K}_1$ into a single kernel, $K_{f_1}$ (named "Full Fusion"). As there is only one kernel in $\mathsf{K}_2$, it will not require any fusion. For the sake of comparison, we created a non-optimal fusion, where we fused $K_1, K_2$ into $K_{f_1^2}$ and $K_3, K_4, K_5$ into $K_{f_2^2}$. Both of these are referred as "Two Fusion". In order to compare GPU performance, we also executed the algorithms in the corresponding CPUs. We'll refer these as "CPU" executions.

All these algorithms in Table II are general enough to be considered as representative algorithms in the image processing domain. Our experiments have shown that kernel fusion is indeed useful. We also calculated the amount of data transfers, data utilization and others for "No Fusion", "Two Fusion" and "Full Fusion". We bench-marked the total execution time for different data box sizes while executing "No Fusion" and "Full Fusion" kernels. The results obtained were used to calculate the speed up for various fusion options.

Table IV: Dependency Types of Kernels

| Algorithms | Kernel Name | Dependency Type |
|---|---|---|
| Convert RGBA to Gray | $K_1$ | Thread to Thread |
| IIR Filter | $K_2$ | Thread to Thread |
| Gaussian Smooth Filter | $K_3$ | Thread to Multi-thread |
| Gradient Filer | $K_4$ | Thread to Multi-thread |
| Threshold Computation | $K_5$ | Thread to Thread |
| Apply Kalman Filter | $K_6$ | Kernel to Kernel |

We considered the following restrictions while generating the compact kernel, a) the given order of the kernels cannot be violated, b) execution of one kernel ($K_i$) starts after finishing the execution of the previous one ($K_{i-1}$), c) the amount of shared memory each one of the kernels and the resulting $K_f$ are using, can not be greater than the total allocated shared memory size, d) all the kernels will be computing on the same input data, e) no thread in the fused kernels will depend on any thread from other blocks, f) no block will depend on threads from other blocks. It was also reasonable to assume that, all the previously known CUDA kernel optimization techniques were implemented for executing the kernels.

## VIII. RESULT ANALYSIS

In this section, we will discuss our findings and explain them. All of the measured times will be in $msec$, unless otherwise mentioned.

We have executed both simple and fused kernels on different devices. Fig 15a is showing execution time for simple and fused kernels. We have computed for three different input sizes, by varying their spatial dimensions. For each of the input size, we benchmarked the execution time for different data box dimensions ($Box_b$ has ($x \times y \times t$) dimensions). The spatial dimensions ($x, y$) of each of the boxes were also varied as (16 × 16), (32 × 32), and (64×64). The temporal dimension for simple kernels were, $t = 1$. The temporal dimension for the fused kernels were calculated using eq (6). From the figure, it is clear that, for each input size and for each data window size, fused kernels performed better than simple kernels. An increase in input size, increases the execution time.

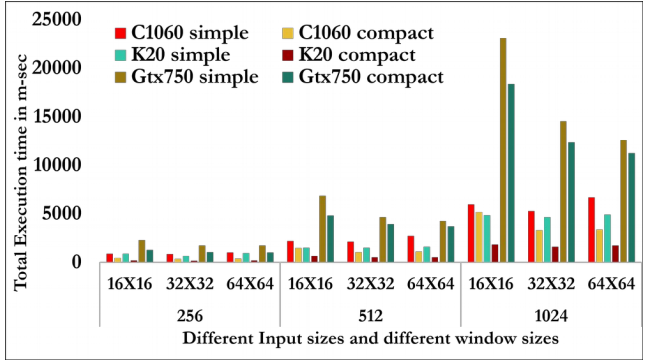

Figure 9: Simple vs Fused Kernel Execution Times for Different Input Dimensions in Different Devices. The x-axis representing image dimensions (256 × 256, 512 × 512, 1024×1024) For each of the input, we bench-marked the execution time for different sizes of data box. Here, (16 × 16), (32 × 32), (64 × 64), representing the spatial dimensions $x, y$ of a box $Box_{b_{in}}$. For simple kernels, the temporal dimension was, $t = 1$ and for fused kernels $t$ is calculated by equation (6). The y-axis is showing total execution time for computing $B = \frac{NMT}{xyt}$ blocks

Fig 10 is comparing the best and worst case GPU timing with the serial execution in the respective CPUs. GPU worst case execution times were corresponding to the computation timing for non-optimal resource allocation for simple kernels. On the contrast, the best timings are corresponding to the execution timings for optimal data and resource allocation for fused kernels. In both these two cases, the spatial dimensions were 32 × 32 The CPU times were collected by executing the serial process in the CPUs, hosting the respective CUDA enabled GPUs.

Fig 11 is showing the speed up achieved by fused kernels w.r.t to serial and sequential processes. Fig 11a is the speed up achieved by fused kernel w.r.t. to serial process. Similarly, a fused kernel will gain speed up in contrast to the sequential kernel execution. Fig 11b is showing speed up achieved by fused kernels in different devices, in contrast to its sequential counterpart. Here, we compared the speedup, for different input sizes and variations in data box sizes.

We showed in Fig 7 the varying data utilization for different devices, when we are changing the box dimensions. Higher data utilization eventually leads to reduced total data movement. Fig 12 comparing pixel transfer and data utilization. Fig 12a comparing total pixel transfers for a given input of dimension 256 × 256 × 1000

Fig 12a is showing number of pixel transferred for different box sizes, when we execute 5 kernels sequentially, two fused kernels and full fusion. In case of no fusion, the number of data movement is constant. For two fusion and full fusion, we see a variation in transfers, for changes in box sizes. Among the data box sizes, for [8, 8, 8]size, two kernel has shown worse performance than no fusion. But for other cases, the data transfers were decreased gradually for an increase in box size.

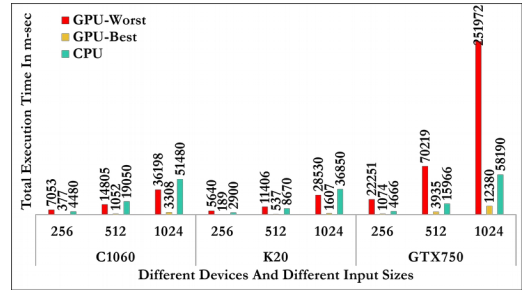

Figure 10: GPU vs CPU Execution Times Comparison. GPU-Best time is the best time found for CuKer. GPU-worst time is the simple kernel execution time for minimum thread allocation. CPU time is representing the serial execution time in the respective host CPUs.

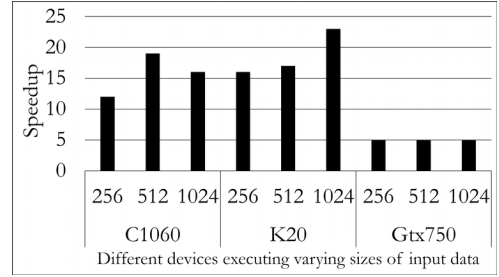

(a) Fused vs Serial Process

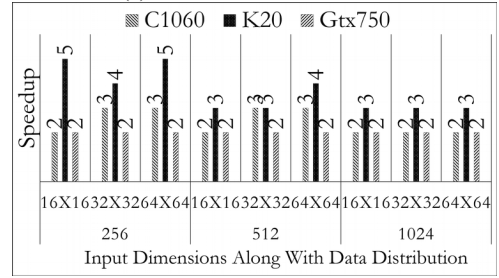

(b) Fused vs Simple Kernels

Figure 11: Speed up of Fused kernel in comparison with Simple Kernels and Serial Process.

Fig 12b is showing precentage reduction in data movement for two fusion and full fusion. It also shows the data utilization values corresponding to the data movement reduction. It is evident that, the amount of reduction in data movement is highly correlated with data utilization. Usage of GMEM in case of "No Fusion", "Two Fusion", and "Full Fusion" are shown in Fig 13. For different sizes of input data, both "Two Fusion" and "Full Fusion" has reduced GMEM usage (33% and 44%, respectively).

Fig 15 is showing nvprof profile timing diagram for simple and fused kernel execution. We used **nvprof** for generating the images for K20 device. The box sizes for this experiment were $[32 \times 32 \times 16]$ and $[32 \times 32 \times 1]$ for fused and simple kernels respectively. Fig 15a is showing the execution of fused kernel, which is calculating 16 frames in this particular example. In contrast, Fig 15b is showing the execution of single kernels, in a sequence, calculating 1 frame. Time required for fused kernels (for 16 frames) is $\approx 490\mu sec$, that is, $31\mu sec$ per frame, where as time required to process 1 frame by simple kernel is $\approx 64\mu sec$. Computation time for 16 frames will include system overhead time. In reality, we found it took $\approx 1400\mu sec$ to finish operations on 16 frames.

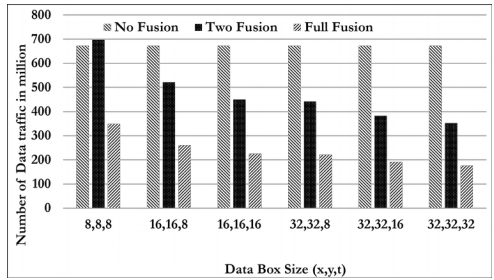

(a) Comparison of Pixel Transfers

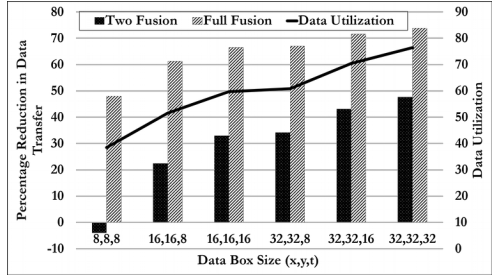

(b) Pixel Transfer and Data Utilization

Figure 12: Comparison of Reduction in data movement and Data Utilization. We generated this figure for input size of $256 \times 256 \times 1000$. Number of data movement, percentage reduction for fused kernels, and data utilization values in different devices will be the same. Change in box dimensions caused the variations of the values.

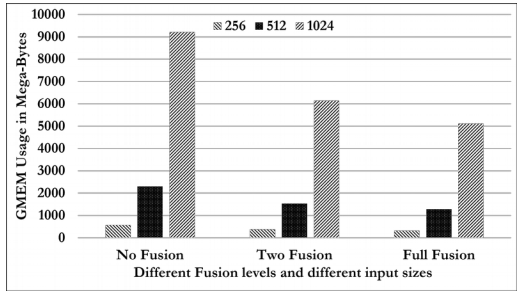

Figure 13: Comparison of GMEM memory usage. No Fusion uses maximum GMEM. Full Fusion enables to use less GMEM. Two Fusion and Full Fusion enables to reduce 33% and 44% GMEM usage.

Fig 14 is showing calculated throughput achieved by simple and fused kernels, for different input sizes executed in different devices. In our input cases, we varied the spatial dimensions, and kept the temporal dimensions same. An increase in input dimension impose extra pressure on computation. It is evident form the figure that, application of kernel fusion would definitely enable us to process more frames in comparison to sequential process.

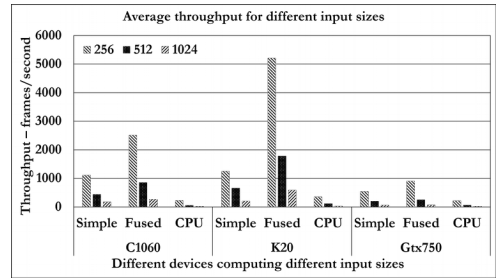

Figure 14: Calculated throughput in terms of Frames/Second for different devices computing on different input size. As we are interested to compute HSDV segments in near-real time, it is important to identify throughput in term of Frames/Second.

IX. RELATED STUDY

CUDA enabled GPU devices have assisted domain experts in developing faster solutions exploiting parallelism for relevant problems. For designing a CUDA kernel code, developers have to iterate through the following steps: a) Develop, b) Parallelize, c) Optimize and d) Deploy. Optimization took substantial part of developing an efficient CUDA kernel. Researchers have been trying to provide techniques for developing such code.

GPU performance improvement via kernel fusion is introduced in many contexts. Guibin et. al [7] proposed kernel fusion technique for energy efficiency. The research was only focusing on reducing the usage of hardware resources, by fusing a small number of kernels. Filipovic et. al [8] showed a library based kernel fusion. Their technique is to generate meta-information of BLAS (basic linear algebra subroutine) functions, represent kernel operations as a high level function. So, when multiple kernels are represented by using those basic functions, an analyzer can analyze the code in meta-information, re-generate new sets of kernels. Fousek et. al [9] also showed a similar map-reduce based performance improvement using kernel fusion. Haicheng et. al [10,11] proposed automatic kernel fusion technique, based on similar principle for Data Base query operations using CUDA device.

In a very recent paper, Wahib et. al [6] presented kernel fusion method for fusing multiple kernels accessing different arrays. In their problem, they initially grouped multiple kernels together based on their data dependency, and later fused the kernels in the same group. Group of kernels were flexible about the sequence of execution. They represented the kernel fusion problem as an optimization problem and proposed a method to solve the problem using their performance prediction method. One of the limitations of their model was, the fused kernels already existing in code base. Also, the fused kernels were not

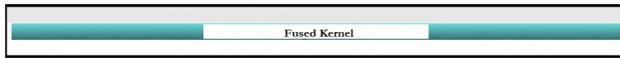 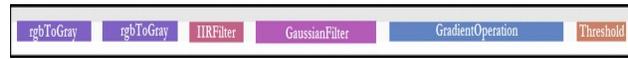

(a) Compact Kernel(CuKer), Executing on 16 frames. Each thread block is computing a data box of size, $Box_{b_{in}} = [32, 32, 16]$

(b) Individual Kernels, Executing on a single frame. Each thread block is computing a data box of size, $Box_{b_{in}} = [32, 32, 1]$

Figure 15: Simple and Fused kernel execution timing diagram generated by nvprof profile

restricted to be executed in a strict sequence. For computing multi-dimensional arrays, their model did not mentioned the data distribution policy for efficient performance.

Image processing domain lacks the existence of set of basic operators or operations, which is universally accepted. As a result, it would be very challenging to represent a given image processing algorithm as combination of basic operators. Also, different algorithms have different data dependency. So, for image processing domain, there are a few major challenges, a) identify the partition of given kernels, b) compose efficient kernels, c) efficiently distribute input data and d) allocate resource, in order to optimally execute the fused kernel(s). Another challenge is, algorithms might be SIMD in nature, but, they have to be executed in a restricted sequence.

The kernel fusion technique is significantly different from kernel tuning technique. Techniques for improving the performance of a single kernel were developed by different researchers [12]–[14]. Application of these techniques for kernel development were shown to improve performance for target GPU devices. Kernel tuning techniques, manages data and resources exclusively used by the kernel. In the presence of multiple kernels executing on multidimensional data, the overall data utilization depends on the data access patterns by the algorithms. So, for multiple kernel executing on various number of data arrays, it is required to develop new technique to manage data movement among the kernels.

## X. CONCLUSION

In this paper we studied application of kernel fusion for analyzing large volume of video data. Our fusible kernel identification method is very easy to implement. The kernel fusion algorithm fused kernel, which will reduce data traffic and reduce GMEM usage. The data distribution algorithm further ensures the reduction in data traffic. Overall, our method considers, data access patterns imposed by different algorithms, and available resource parameters to achieve minimal total execution time. Empirical observations verified the effectiveness of the proposed method. We presented the evaluation of the model using a limited set of algorithms. But, the algorithms are representative algorithms form the domain, and our methods can be used for extended set of such algorithms. Our future goal is to develop a set of basic algorithms, which can be used to represent any IPA. This will lead to develop automated fused kernels, in contrast to our manual process.

## REFERENCES

[1] T. Luhmann, "Close range photogrammetry for industrial applications," {ISPRS} Journal of Photogrammetry and Remote Sensing, vol. 65, no. 6, pp. 558 – 569, 2010. {ISPRS} Centenary Celebration Issue.

[2] E. D. Ross and V. K. Pulusu, "Posed versus spontaneous facial expressions are modulated by opposite cerebral hemisperes," Cortex, vol. 45, no. 5, pp. 1280–1291, 2013.

[3] J. Kimbell, E. Gross, D. Joyner, M. Godo, and K. Morgan, "Application of computational fluid dynamics to regional dosimetry of inhaled chemicals in the upper respiratory tract of the rat," Toxicology and Applied Pharmacology, vol. 121, no. 2, pp. 253 – 263, 1993.

[4] http://www.gurobi.com/, "Gurobi optimizer."

[5] "CUDA Programming Guide." http://docs.nvidia.com/cuda/cuda-c-programming-guide/.

[6] M. Wahib and N. Maruyama, "Scalable kernel fusion for memory-bound gpu applications," in Proceedings of the International Conference for High Performance Computing, Networking, Storage and Analysis, SC '14, pp. 191–202, 2014.

[7] G. Wang, Y. Lin, and W. Yi, "Kernel fusion: An effective method for better power efficiency on multithreaded gpu," in Proceedings of the 2010 IEEE/ACM Int'L Conference on Green Computing, pp. 344–350, 2010.

[8] J. Filipovic, M. Madzin, J. Fousek, and L. Matyska, "Optimizing CUDA code by kernel fusion—application on BLAS," CoRR, vol. abs/1305.1183, 2013.

[9] J. Fousek, J. Filipovič, and M. Madzin, "Automatic fusions of cuda-gpu kernels for parallel map," SIGARCH Comput. Archit. News, vol. 39, pp. 98–99, Dec. 2011.

[10] H. Wu, G. Diamos, S. Cadambi, and S. Yalamanchili, "Kernel weaver: Automatically fusing database primitives for efficient gpu computation," in Proceedings of the 2012 45th Annual IEEE/ACM International Symposium on Microarchitecture, MICRO-45, (Washington, DC, USA), pp. 107–118, IEEE Computer Society, 2012.

[11] H. Wu, G. Diamos, J. Wang, S. Cadambi, S. Yalamanchili, and S. Chakradhar, "Optimizing data warehousing applications for gpus using kernel fusion/fission," in Proceedings of the 2012 IEEE 26th International Parallel and Distributed Processing Symposium Workshops, pp. 2433–2442, 2012.

[12] A. Davidson and J. D. Owens, "Toward techniques for auto tuning gpu algorithms," Para 2010: State of the Art in Scientific and Parallel Computing, June 2010.

[13] S. Grauer-Gray, L. Xu, R. Searles, S. Ayalasomayajula, and J. Cavazos, "Auto-tuning a high-level language targeted to gpu codes," in Innovative Parallel Computing (InPar), 2012, pp. 1–10, May 2012.

[14] J. A. Stratton, N. Anssari, C. Rodrigues, N. O. T, L. Chang, and G. D. Liu, "Optimization and architecture effects on gpu computing workload performance," Innovative Parallel Computing, pp. 1–10, 2012.